%% file: Main.tex
\documentclass[conference]{IEEEtran}
\usepackage[pdftex]{graphicx}
\usepackage[usenames,dvipsnames,table,xcdraw]{xcolor}
\usepackage{array}
\usepackage{multirow}
\usepackage[normalem]{ulem}
\useunder{\uline}{\ul}{}
\usepackage{booktabs}
\usepackage{supertabular,booktabs}
\usepackage{longtable}
\usepackage{lscape}
\usepackage{cite}
\usepackage{url}
\usepackage{hyperref}
\usepackage{dirtytalk}
\usepackage{graphicx}
\usepackage{comment}
\usepackage{makecell}
\usepackage{svg}
\usepackage{amsmath}

\usepackage[caption=false]{subfig}
\graphicspath{{Images/}}
\usepackage[normalem]{ulem}
\useunder{\uline}{\ul}{}

\usepackage{tabularx}
\usepackage{balance}
\usepackage{nameref}

\newcommand{\MyBox}[1]{\vspace{3mm}\noindent\framebox[\columnwidth][c]{\parbox[b]{0.95\columnwidth}{ #1 }}\vspace{3mm}}

\begin{document}

\title{Diversity in Software Engineering Education: Exploring Motivations, Influences, and Role Models Among Undergraduate Students \\
}

\author{
\IEEEauthorblockN{Ronnie de Souza Santos}
\IEEEauthorblockA{University of Calgary\\
Calgary, AB, Canada \\
ronnie.desouzasantos@ucalgary.ca} 
\and

\IEEEauthorblockN{Italo Santos}
\IEEEauthorblockA{Northern Arizona University\\
   Flagstaff, AZ, US \\
   ids37@nau.edu}
\and

\IEEEauthorblockN{Robson Santos}
\IEEEauthorblockA{UFPE\\
   Recife, PE, Brazil \\
  rtss@cin.ufpe.br}
\and

\IEEEauthorblockN{Cleyton Magalhaes}
\IEEEauthorblockA{UFRPE\\
 Recife, PE, Brazil \\
  cleyton.vanut@ufrpe.br}
}

\maketitle

\begin{abstract}
    Software engineering (SE) faces significant diversity challenges in both academia and industry, with underrepresented students encountering hostile environments, limited representation, and systemic biases that hinder their academic and professional success.
    Despite significant research on the exclusion experienced by students from underrepresented groups in SE education, there is limited understanding of the specific motivations, influences, and role models that drive underrepresented students to pursue and persist in the field.
    This study explores the motivations and influences shaping the career aspirations of students from underrepresented groups in SE, and it investigates how role models and mentorship impact their decisions to stay in the field.
    We conducted a cross-sectional survey with undergraduate SE students and related fields, focusing on their motivations, influences, and the impact of mentorship and role models on their career paths.
    We identified eight motivations for pursuing SE, with career advancement, technological enthusiasm, and personal growth being the most common. Family members, tech influencers, teachers, and friends were key influences, though 64\% of students reported no specific individual influence. Role models, particularly tech influencers and family members play a critical role in sustaining interest in the field, especially for underrepresented groups.
    This study provides insights into the varied motivations and influences that guide underrepresented students' decisions to pursue SE. It emphasizes the importance of role models and highlights the need for intersectional approaches to better support diversity in the field.
\end{abstract}


\section{Introduction}

Technology shapes many aspects of contemporary life, influencing work, education, politics, and leisure. Despite the multicultural and diverse nature of modern society, software engineering (SE) is central to technological progress, and it struggles with a notable lack of diversity in both academic and industrial settings. This lack of diversity and inclusivity presents significant challenges in creating software solutions that truly reflect and serve the needs of our society \cite{albusays2021diversity, silveira2019systematic, rodriguez2021perceived}. 

Gender disparity in SE exemplifies the consequences of the field's diversity deficit \cite{wang2017diversity, sax2017diversifying}. Toxic cultures characterized by hostility and sexism often prevail in software development environments, creating unwelcoming spaces for women \cite{zolduoarrati2021value}. These environments frequently foster gender-based discrimination through unequal opportunities and biased evaluations, further inhibiting women's professional growth. These systemic challenges impede career advancement and reinforce a culture of exclusion, ultimately diminishing diversity and inclusion \cite{oliveira2023navigating}.

In SE education, the situation is similarly challenging. Students from underrepresented groups frequently encounter hostile and unwelcoming environments within SE programs, which hinder their sense of belonging and impede academic success \cite{ibe2018reflections, ouhbi2020software}. Contributing factors include the lack of representation and visibility of these groups in the curriculum, limited support networks, and biases present in course content and classroom dynamics. Moreover, instances of implicit or explicit discrimination—whether based on gender, race, sexual orientation, or other identities—further create hostile learning environments \cite{ibe2018reflections, sax2017diversifying, cheryan2013stereotypical, de2023lgbtqia+}.

In particular, women in SE education often face isolation and exclusion, exacerbated by the male-dominated nature of the field. This imbalance fosters a lack of peer support and mentorship, which is critical for academic and professional growth \cite{oliveira2023navigating}. Additionally, women are frequently subject to stereotypical expectations that question their technical competence and undermine their contributions \cite{cheryan2013stereotypical}. These challenges are compounded by gender biases in assessments and interactions with faculty, further reinforcing their underrepresentation and diminishing their confidence in pursuing long-term careers in the field \cite{silva2024paths}. Consequently, many women feel discouraged and may ultimately leave the program or shift to other disciplines, perpetuating the gender gap in SE \cite{oliveira2023navigating, sax2017diversifying, silva2024paths}.

Considering the persistent lack of diversity awareness as a continuous pipeline from academia to industry, alongside numerous studies highlighting the exclusion experienced by students from underrepresented groups in SE education \cite{zolduoarrati2021value, richard2022lgbtq}, our research focused on exploring the influences and motivations that guide students' decisions to pursue careers in SE and related fields. This approach allowed us to investigate \textbf{RQ. What key factors shape students' professional aspirations and understand how these dynamics interact with the challenges faced by underrepresented groups?} Specifically, we aim to answer the following research questions:

\begin{description} \item[RQ1] What are the key motivations driving students from underrepresented groups to pursue careers in software engineering and related fields? 

\item[RQ2] What influences, both personal and external, shape the career decisions of these students? 

\item[RQ3] How do role models and mentorship impact the career aspirations of students from underrepresented groups in software engineering? \end{description}

To address our research question, we conducted a cross-sectional survey to examine the motivations, influences, and role models of undergraduate students enrolled in SE and related fields. This paper offers four key contributions: (i) identified eight motivations for pursuing SE, with career advancement, technological enthusiasm, and personal growth being the most common, and highlighted variations across gender, ethnicity, and sexual orientation; (ii) emphasized the role of family, tech influencers, teachers, and friends as key influencers, while noting that 64\% of students reported no specific influence, suggesting independent career choices; (iii) demonstrated the importance of long-term role models, particularly for underrepresented groups; and (iv) provided intersectional insights into how motivations, influences, and role models differ across diverse demographic groups.


The structure of this paper is organized as follows. Section 2 provides a background on research related to SE education and underrepresented groups. Section 3 outlines our methodology based on a global survey of students. In Section 4, we present the survey results. Section 5 discusses the findings and their broader implications. Finally, Section 6 concludes the study, summarizing key insights and suggesting potential future directions.

\section{Background and related work}

Although students’ interest in science often diminishes as they advance through their education~\cite{osborne2003attitudes}, the learning environments and instruction they encounter can have a positive influence on their motivational beliefs and long-term persistence~\cite{chittum2017effects, fortus2014measuring}. Leaper and Starr~\cite{leaper2019helping} argue that gender inequality in science, technology, engineering, and mathematics (STEM) is primarily driven by motivation rather than academic performance disparities. It is essential to inspire all students, regardless of gender, to engage with STEM subjects and motivate many to pursue careers in these fields~\cite{dokme2022investigation, pcast2010k}. As Simpkins et al.~\cite{shmueli2008child} emphasized, every available resource should be leveraged to encourage students to develop academic preferences that guide their career choices in STEM fields.

While many factors contribute to motivating students to pursue STEM and persevere in these fields, the teacher's role in fostering social motivation is considered crucial~\cite{williams2019will}. Understanding how this factor influences students' motivation is dependent on various elements, such as the teacher’s characteristics, beliefs, and teaching methods~\cite{wentzel2009students}. Role models are instrumental in promoting inclusion and a sense of belonging across disciplines, serving as tangible examples of the goals students strive to achieve~\cite{doherty2016role, rosenthal2013pursuit}. These role models are often relatable figures whose success aligns with students' ambitions~\cite{morgenroth2015motivational}. In STEM fields, studies have demonstrated the impact of role models on efforts to diversify these areas~\cite{gladstone2021role}. Similarly, across different disciplines, influential figures from underrepresented groups play a vital role in inspiring students from marginalized backgrounds, helping them build resilience, stay engaged, and persist in their academic pursuits~\cite{susor2018role, karunanayake2004relationship}.

Role modeling as a teaching strategy has long been recognized across numerous disciplines as an effective way to empower students and enhance the development of their professional skills~\cite{key2021role, hu2020not}. Historical figures play a unique role in this approach, as their remarkable achievements and resilience offer timeless lessons that continue to inspire learners in modern society~\cite{givens2016modeling}. In fields like SE, where diversity is often lacking, it is particularly important for students, especially those from marginalized backgrounds, to move beyond stereotypical portrayals of scientists in media and textbooks. Exposing these students to renowned scientists from underrepresented groups, who have made significant contributions to technology, can provide them with relatable role models and inspire their aspirations~\cite{nelson2022role}. 

Moreover, by understanding the motivations and inspirations that guide students' choices to pursue a career in STEM and the impact of relatable role models, we can create more inclusive environments. Promoting diversity in these areas is essential, as diverse role models help foster a sense of belonging and inspire a wider range of students to pursue careers in these fields, ultimately contributing to a more equitable and innovative workforce. Our study examines the motivations and factors influencing the career aspirations of students from underrepresented groups in SE, focusing on how role models and mentorship shape their decisions to pursue and remain in the field.

\section{Method}\label{method}
We conducted a cross-sectional survey \cite {easterbrook2008selecting, pfleeger2001principles, linaker2015guidelines} in accordance with established guidelines in SE research \cite {ralph2020empirical} to explore the motivations, influences, and role models of undergraduate students enrolled in SE and related courses, such as computer science, and information systems. In this process, we adopted a hybrid data collection strategy by integrating referral sampling, snowball sampling, and using Prolific \footnote{www.prolific.com} as a platform to distribute the questionnaire to students worldwide. This approach allowed us to reach a diverse and global sample of students. We then applied quantitative and qualitative analysis techniques to synthesize the data collected, enabling us to identify patterns and insights across gender, ethnicity, and sexual orientation. Below, we describe this methodology in detail.

\subsection{Questionnaire}\label{questionnaire}
We designed an anonymous questionnaire for data collection, comprising four (4) main sections, each aimed at exploring different aspects of participants' backgrounds and experiences in SE:

\begin{itemize} \item \textit {Section I: Motivations}: This section focused on collecting data on why participants decided to pursue a career in SE. The aim was to understand the factors that contributed to their career choice, including personal, academic, or professional motivations.

\item \textit {Section II: Influences on Career Decisions}: This section explored the various external and internal influences that shaped participants' decisions to enter the field of SE. The goal was to gather information on how different agents, such as people or experiences, contributed to their career path.

\item \textit {Section III: Role Models}: The purpose of this section was to gather insights about the role models participants identified as influential in their pursuit of SE. This section aimed to explore the extent to which mentors, public figures, or other individuals impacted their career aspirations.

\item \textit {Section IV: Demographic Information}: The final section was designed to collect demographic data, such as participants' country, gender, ethnicity, and other relevant personal information. This section provided the context necessary to analyze the data across different backgrounds and identities. \end{itemize}

\input{tables/survey}

Since our study emphasizes diversity and inclusion, understanding how students from underrepresented groups relate to various aspects of their choices to pursue a career is essential. Therefore, we explicitly asked participants about their gender, ethnic group, and sexual orientation while ensuring that they had the option to refrain from answering if they were not comfortable sharing these personal details, e.g., questions related to sexual orientation and belonging to the LGBTQIA+ community were optional.

\subsection{Population} \label{sec:participants}
Determining the global population of SE students is complex due to variations in the structure and definition of programs across different institutions and countries. Some universities offer dedicated degrees in SE, while others integrate SE into broader fields like computer science or engineering. Additionally, the terminology and scope of these programs can vary, with some emphasizing software development while others treat it as a smaller component of a more general technical education. These differences make it difficult to establish a uniform definition of who constitutes a ``SE student", further complicating efforts to capture this population globally.

Moreover, practical challenges such as limited access to specific student groups, language barriers, and uneven participation across countries make achieving a fully representative sample even more difficult. Certain regions or institutions may have been underrepresented due to logistical hurdles or differences in outreach. As a result, our study targeted undergraduate students from various countries enrolled in programs that included some level of SE, whether as a dedicated degree or a module within another discipline. Therefore, in our study, the population included undergraduate students from different countries enrolled in university programs that incorporate SE to some degree and who were willing to participate.

\subsection{Sampling and Data Collection} \label{sec:collect}
We utilized a combination of non-probability sampling methods to gather data for our study. Initially, following the protocol from the original study we are replicating, we began with convenience sampling by reaching out to SE professors from universities around the world. These professors were identified through their involvement in various SE conference program committees, which provided a reliable and accessible pool of contacts within the field. We formally requested their assistance in distributing our questionnaire to undergraduate SE students through email lists, learning management systems, or any other communication channels available at their institutions. This method helped us to quickly reach students already connected to researchers in the SE academic community, though it was naturally limited by the networks of the professors contacted.

To broaden our reach beyond these initial contacts, we incorporated snowball sampling, where participants were encouraged to share the questionnaire with their peers and fellow students who might also meet the study’s criteria. This method enabled the study to grow organically as students passed the survey on to other potential participants, increasing the study’s reach in a cost-effective manner. Snowball sampling also allowed us to engage with students who may not have been directly accessible through the initial convenience sampling phase, helping to diversify the sample in terms of geography, educational backgrounds, and academic experiences.

In addition to convenience and snowball sampling, we employed the Prolific platform to further support participant recruitment and expand the pool beyond academic networks. Prolific is a widely utilized crowdsourcing platform, boasting over 150,000 active users \cite {russo2022recruiting}, and has been extensively used for conducting SE surveys across various contexts. This platform allowed us to engage a broader spectrum of participants from multiple geographical regions, contributing to a more representative sample. Prolific’s large user base offered access to students beyond the immediate academic contacts, facilitating international participation and helping us overcome regional limitations inherent in traditional academic recruitment methods.

However, using Prolific also introduced the potential for including participants outside our target population of SE students. To mitigate this, we implemented multiple screening mechanisms within our data collection process. A key component was a pre-screening validation section that incorporated several filters and skill-based screening questions designed to verify the participants' background in SE. These questions were based on recommended assessments of basic programming skills \cite {danilova2021you}, including fundamental knowledge such as variable types and a competency expected of students in the field. Only participants who successfully passed this pre-screening validation section were granted access to the full survey. This ensured that only individuals with relevant educational backgrounds participated in the study, preventing inaccurate or random responses that could affect the study's validity.

Data collection was concluded with 89 participants from 20 different countries. These participants were studying in programs that included some level of focus on SE and represented a broad array of cultural and academic backgrounds. This combination of sampling methods, together with the filtering mechanisms, allowed us to gather diverse and reliable data while managing the inherent limitations associated with non-probability sampling techniques.

\subsection{Data Analysis} \label{sec:analyse}
In this study, both quantitative and qualitative data were collected. However, qualitative data was mostly quantifiable. In this study, both quantitative and qualitative data were collected. However, qualitative data was mostly quantifiable. Hence, we used thematic analysis \cite{cruzes2011recommended} to convert qualitative data from open-ended survey questions into categories, allowing for quantitative analysis. After reviewing and coding the responses based on recurring themes such as motivations, influences, and role models, we categorized the data into structured groups. Hence, we employed descriptive statistics \cite {george2018descriptive, de2019evolution} to explore the primary characteristics of our sample and answer the research questions. 

Descriptive statistics is a method used to summarize and describe the essential features, characteristics, and patterns within a dataset. This approach encompasses measures like central tendency (such as mean, median, and mode), variability (e.g., range, variance, standard deviation), and distribution representations (such as histograms and frequency distributions) \cite {george2018descriptive}. In survey analysis, descriptive statistics are a practical approach for obtaining a structured overview of the collected information.

Our data analysis process involved segmenting participants' responses into subgroups using statistical measures like means, proportions, totals, and ratios to gain a thorough understanding of the answers obtained. We conducted this process supported by Tableau \footnote{https://www.tableau.com/}, which allowed us to create interactive visualizations that aided in analyzing data patterns such as averages and distributions. Additionally, through aggregations, we explored the correlation between students' backgrounds, experiences, and knowledge of computer scientists from underrepresented groups.

\subsection{Ethics}
Following ethical guidelines, no personal information regarding the participants (such as names, emails, or universities) was gathered during this study to uphold participant anonymity. Before starting the questionnaire, participants were briefed about the study's objectives and were required to consent to use their responses for scientific purposes.

\section{Findings}
As detailed in Section~\ref{sec:collect}, we employed a combination of convenience sampling, snowball sampling, and crowdsourcing to collect data from a broad and diverse population of SE students globally. Initially, we reached out to SE professors to distribute the survey to students within their networks (convenience sampling), while also encouraging participants to share the survey with their peers (snowball sampling). To expand our reach beyond these academic networks, we utilized the Prolific platform for crowdsourcing, which provided access to a broader demographic range. This multifaceted approach resulted in a participant pool of 89 individuals from 20 countries, ensuring a geographically diverse sample.

Our sample also reflects diversity across several key demographic dimensions, including the variety of SE programs students were enrolled in, their academic year of study, and the motivations influencing their decision to pursue a career in SE. This diversity is important for the study, as it enables us to examine how students from different educational and cultural backgrounds perceive the contributions of scientists from underrepresented groups in the history of SE. Additionally, the sample captures diversity in gender, ethnicity, and sexual orientation, which is essential to our focus on understanding the experiences of students from traditionally underrepresented backgrounds in STEM fields. A summary of the demographic breakdown of our sample is provided in Table \ref{tab:Demographics}.

\input{tables/Demographics}

\subsection{Motivations for Students Pursuing a Career in SE}

In our study, we identified eight distinct motivations for students pursuing a career in SE. \textit{Career Advancement} reflects students' aspirations to secure stable, well-paying jobs with opportunities for professional growth, promotions, and long-term career prospects. \textit{Technological Enthusiasm} captures a genuine interest in the field, where students are by curiosity and a desire to engage with cutting-edge developments. \textit{Personal Growth} involves students who view SE as a way to challenge themselves, develop their problem-solving skills, and gain expertise. \textit{Early Age Contact} refers to the motivation arising from exposure to technology at a young age—whether through computers, programming, or other forms of digital interaction. \textit{Personal Interest} is driven by an intrinsic interest in the field itself, without necessarily being tied to the excitement of technological innovation or self-improvement. \textit{Previous Training} refers to the motivation resulting from formal or informal education in technology or programming before entering their current studies. \textit{Social Contribution} is a desire to use technology to solve societal and world challenges. Finally, \textit{Parental Influence} is a motivation shaped by the guidance, encouragement, or expectations of their parents or guardians.

Considering the entire group of 89 students, the most common motivation for pursuing a career in SE was \textit{career advancement}, cited by \textit{30\%} of participants (27 out of 89). This was followed by \textit{technological enthusiasm}, which motivated \textit{20\%} of the students (18 out of 89), and \textit{personal growth}, cited by \textit{15\%} of the participants (14 out of 89). A smaller percentage of students indicated \textit{early-age contact} with technology (\textit{11\%} or 10 students), while \textit{personal interest} also motivated \textit{11\%} (10 students). \textit{Previous training} accounted for \textit{6\%} of the motivations (5 students), while \textit{3\%} (3 students) were motivated by \textit{social contribution}. \textit{Parental influence} was the least common motivator, reported by only \textit{2\%} (2 students).

Across all \textit{gender groups}, career advancement was the most common motivation for pursuing a career in SE, reflecting a strong focus on professional growth and stable career opportunities for those interested in pursuing this area. \textit{Male students} who cited this motivation most frequently also displayed technological enthusiasm as a strong motivator, indicating a fascination with the evolving nature of technology. In contrast, \textit{female students}, while also prioritizing career advancement, were more driven to SE and motivated by personal growth and their early contact with technology, which highlights the importance of self-improvement and early exposure to technology in shaping their career decisions. \textit{Non-binary students}, though only two in the sample, showed an equal emphasis on career advancement and personal interest, suggesting a more personalized approach to their career choices, but this observation requires more data for further exploration. Overall, while career advancement is a unifying factor, motivations such as technological enthusiasm, personal growth, and early exposure to technology vary in significance across different genders.

Across all ethnicities, career advancement emerged as the dominant motivation. Among \textit{Asian/Pacific Islander} students, it was the most common motivator, followed by personal growth. \textit{Black/African American} students displayed a more varied set of motivations, with both career advancement and technological enthusiasm cited equally. Notably, social contribution was a significant motivator for this group, making them the only ethnic group in the sample that cited the societal impact of technology as a motivation. \textit{Hispanic/Latinx} students demonstrated a balance between early-age contact with technology and personal growth, while motivations like technological enthusiasm, personal interest, and previous training followed closely. For \textit{Multiracial/Mixed Ethnicity} students, motivations were evenly split between career advancement and early-age contact. Finally, among \textit{White/Caucasian} students, career advancement was again the leading motivator, followed closely by technological enthusiasm and personal growth. However, as the majority group in the sample, other motivators were also observed, with students citing personal interest, early-age contact, and previous training as key factors in their decision to pursue SE.

Among \textit{heterosexual students}, career advancement was the dominant motivation, cited by a significant portion of participants. Other important motivations included technological enthusiasm, personal growth, early-age contact with technology, technology, and motivating factors. On the other hand, \textit{LGBTQIA+ students} also prioritized career advancement when choosing to follow SE, though technological enthusiasm is also a key motivation for these students, and other motivations were more evenly distributed across other categories, including personal growth, personal interest, early age contact with technology and previous training as key influences. 

Looking closer to the sexual orientations presented in the sample, among \textit{bisexual} students, \textit{career advancement} was the most frequent motivation, followed closely by \textit{technological enthusiasm}, with some also citing \textit{early age contact} and \textit{previous training} as key influences in their career decisions. \textit{Gay} students displayed a more balanced distribution, with \textit{career advancement}, \textit{personal growth}, \textit{technological enthusiasm}, and \textit{personal interest} all being cited, suggesting that personal fulfillment and a passion for technology play major roles alongside career goals. \textit{Lesbian} students were predominantly motivated by \textit{career advancement}, with both participants identifying it as their primary motivator. The \textit{pansexual} student presented in the sample also cited \textit{career advancement} as their main influence, while the \textit{asexual} student in the sample identified \textit{personal interest} as their primary motivator. 

Finally, considering four students in the sample who identify as \textit{transgender}, career advancement was the most frequent motivator. Additionally, personal growth and technological enthusiasm were each cited by one student, indicating that, while professional development was a key driver, personal fulfillment and enthusiasm for technology also contributed to their decisions to pursue SE. One transgender student also mentioned early age contact as a factor that influenced their career choice. Table \ref{tab:motivations} summarizes the findings of this section.

\MyBox{\textbf{\emph{RQ1 -- Summary.}} \small We identified eight key motivations driving students from underrepresented groups to pursue careers in SE. Career advancement was the most common motivation, followed by technological enthusiasm and personal growth. Motivations varied by gender, ethnicity, and sexual orientation, with male students often driven by technological enthusiasm, while female students prioritized personal growth and early exposure to technology. Black/African American students also highlighted social contribution as a key motivator. LGBTQIA+ students displayed varied motivations, including personal interest, early-age contact, and career advancement.}

\input{tables/motivations}

\subsection{Personal and Professional Influences on SE Career Choice}

Exploring the factors influencing students' decisions to pursue a career in SE, we identified several key agents that played a role in shaping their choices. These agents represent individuals who may have inspired or supported students throughout their decision-making processes. Some students were influenced by \textit{Family Members}, which could include parents, siblings, or other relatives with a background or interest in technology. Others cited \textit{Friends} as key influences, indicating that peers who were already engaged in or enthusiastic about SE may have encouraged them to pursue the field. \textit{Teachers} were also highlighted as significant figures, likely through their guidance, mentorship, or their role in fostering an educational environment that nurtured an interest in the field. For a smaller group, \textit{Tech influencers}—prominent figures in the technology industry, such as innovators, social media influencers, CEOs of tech companies, or public figures known for their contributions to SE—served as a source of inspiration. Additionally, a few students mentioned \textit{Pioneer Scientists}, referring to historical figures in the technology field (e.g., Alan Turing) as pivotal in influencing their career choices. 

In terms of distribution, \textit{64\%} of students (57 out of 89) reported no specific individual as an influence in their decision, indicating that they may have made the choice independently or without the direct encouragement of others. Among those who did cite influences, \textit{tech influencers} accounted for \textit{11\%} (10 students). Additionally, \textit{family members} were reported to have influenced \textit{9\%} (8 students). \textit{Teachers} influenced \textit{7\%} (6 students), and \textit{friends} were cited by \textit{6\%} of students (5 participants), suggesting that peer influence and educational environments played moderate roles. \textit{Pioneer scientists} were mentioned by only \textit{3\%} (3 students), showing that historical figures are not as influential in SE as they might be in other fields.

Across all \textit{gender groups}, \textit{tech influencers} and \textit{family members} were the most common influences on students' decisions to pursue a career in SE, reflecting the role of both personal relationships and public figures in shaping career choices. \textit{Male students} most frequently cited \textit{tech influencers}, followed by \textit{family members} and \textit{friends} as the individuals who influenced them. Notably, they were the only group to mention \textit{pioneer scientists} as an influence, indicating the unique role historical figures play in their decision-making process. In contrast, \textit{female students} were mostly influenced by \textit{family members} and \textit{teachers}, underscoring the importance of supportive personal relationships and educational environments in motivating their pursuit of SE. The sample of \textit{non-binary students} was small, and none of them cited any specific individual as an influence, suggesting that their career decisions might not have been driven by personal or professional figures. Overall, influences on career decisions vary across genders, with males gravitating more towards public figures, while females tend to be influenced by close relationships and mentors.

Considering the reported influences across different ethnic groups, distinct patterns emerged regarding the agents that impacted students' decisions to pursue SE. Among \textit{Asian/Pacific Islander} students, the majority cited \textit{friends} as key influencers, with a smaller number also mentioning \textit{family members} and \textit{teachers}, suggesting a combination of peer and familial influence. For \textit{Black/African American} students, \textit{tech influencers} were the most frequently cited, with additional mentions of \textit{pioneer scientists} and \textit{family members}, reflecting both historical and personal sources of inspiration. \textit{Hispanic/Latinx} students exhibited an equal influence from \textit{tech influencers} and \textit{teachers}, with one participant mentioning a \textit{friend}, indicating a balance between professional and social influences. The small sample of \textit{Multiracial/Mixed Ethnicity} students pointed to both \textit{family members} and \textit{teachers} as their primary influences. Finally, for \textit{White/Caucasian} students, \textit{family members} were the predominant influence, followed by \textit{tech influencers}, with few instances of \textit{friends}, \textit{pioneer scientists}, and \textit{teachers} playing a role, suggesting a more diverse range of influences within this group.

Looking at the reported influences across sexual orientations, some distinct trends emerge. Among \textit{heterosexual} students, \textit{family members} were the most frequently cited influence, followed by \textit{friends}, \textit{tech influencers}, and \textit{teachers}. A small number also mentioned \textit{pioneer scientists}, suggesting that heterosexual students tend to rely on a combination of personal connections and professional figures when deciding to pursue a career in SE. Among the \textit{LGBTQIA+} students, those identifying as \textit{gay} cited \textit{teachers}, \textit{tech influencers}, and \textit{family members} as responsible for influencing their choice. In contrast, \textit{bisexual} students were notably influenced by both \textit{friends} and \textit{teachers}. One student identifying as \textit{lesbian} indicated that a \textit{teacher} influenced her decision. This demonstrates that for LGBTQIA+ students, there is a balance between social and educational environments in shaping their career decisions.

Finally, in our study, only one \textit{transgender} participant cited an influence, which was a \textit{family member}. This lack of cited influences could indicate that transgender students might rely more on personal motivation and internal drivers or perhaps face a lack of visible role models within the field, which could limit their opportunities to identify specific influences in their career choices. Table \ref{tab:influences} summarizes the findings presented in this section.

\MyBox{\textbf{\emph{RQ2 -- Summary.}} \small Our results show that tech influencers and family members were the most common, followed by teachers and friends, while a smaller group mentioned pioneer scientists. Interestingly, 64\% of students reported no specific individual influence, suggesting many made independent career choices. Gender differences emerged: male students were influenced by tech influencers and friends, while female students were more guided by family members and teachers. Non-binary students did not report any specific influences. Ethnic variations showed Asian/Pacific Islander students favored friends, while Black/African American students leaned on tech influencers and pioneer scientists. Hispanic/Latinx students cited a mix of tech influencers and teachers, and White/Caucasian students relied mostly on family members. Heterosexual students primarily cited family members, while LGBTQIA+ students reported more varied influences, including teachers and friends.}

\input{tables/influences}

\subsection{Role Models in SE Education}

When exploring the role models students identified throughout their journey in SE, we observed that these figures were more associated with long-term inspiration rather than immediate influence. While influencers might help guide students toward SE, role models can inspire them to stay in the field and persevere through challenges. These role models offer students aspirational figures to look up to as examples of success and dedication in the field. In our study, some role models fell into the same categories as the influences previously identified.

Some students look up to \textit{Tech Influencers}, including well-known figures in the technology industry, such as CEOs, social media influencers, or innovators who have made significant contributions to SE. Others are inspired by \textit{Pioneer Scientists}, historical figures like Alan Turing or Ada Lovelace, whose foundational work in computing continues to inspire generations. A smaller group is inspired by \textit{Teachers}, who, through their mentorship and guidance, offered valuable examples of dedication to education and the field. Additionally, some students cited \textit{Practitioners}—working professionals in SE—demonstrating that real-world success stories can have a powerful effect on students' aspirations. \textit{Friends} and \textit{Family Members} also serve as role models, providing personal connections to the field and examples of success close to home.

In terms of distribution, the largest group, \textit{61\%} (54 out of 89), did not report any specific role model. For the remaining participants, \textit{Tech Influencers} were the most frequently cited, with \textit{17\%} (15 students) looking up to figures in the technology industry. \textit{Family Members} and \textit{Friends} were each cited by \textit{6\%} (5 students), demonstrating the importance of personal relationships in providing inspiration.. \textit{Pioneer Scientists} were mentioned by \textit{4\%} (4 students), highlighting the importance of historical figures in SE education. Finally, \textit{Practitioners} and \textit{Teachers} were each cited by \textit{3\%} (3 students), showing that educators and professionals in the field also serve as sources of long-term inspiration.

Considering gender, our findings reveal similarities between the influences that initially drew students to SE and the role models that continue to inspire them to stay in the field. Across all \textit{gender groups}, \textit{tech influencers} and \textit{family members} were prominent role models, demonstrating the dual influence of public figures and personal relationships in shaping long-term career engagement. Among \textit{male students}, \textit{tech influencers} emerged as the most frequently cited role models, followed by \textit{family members}, \textit{friends}, \textit{practitioners}, and \textit{pioneer scientists}. Males were the only group to mention \textit{practitioners}, highlighting the unique role of professionals in the field as sources of inspiration. In contrast, \textit{female students} were also revealed to be inspired by \textit{tech influencers}, followed by \textit{pioneer scientists} and \textit{teachers}. Additionally, they cited \textit{family members} and \textit{friends} as important role models, underscoring the value of both personal connections and educational figures in maintaining their interest in the field. For \textit{non-binary students}, the sample was limited, with one citing a \textit{teacher} as a role model. Overall, while role models across genders show slight differences, they are more aligned than when it comes to the influences that initially led students into the field.

Considering ethnicity, our findings show distinct patterns between influences that initially led students into SE and the role models that inspire them to stay. Among \textit{Asian/Pacific Islander} students, the primary role models cited were \textit{tech influencers} and \textit{family members}, with additional mentions of \textit{friends}, \textit{pioneer scientists}, and \textit{teachers}. For \textit{Black/African American} students, \textit{tech influencers} were the most commonly cited role models, followed by \textit{pioneer scientists} and \textit{friends}. \textit{Hispanic/Latinx} students predominantly cited \textit{tech influencers} as their role models. The small sample of \textit{Multiracial/Mixed Ethnicity} students showed only one mention of a \textit{teacher} as a role model. Finally, among \textit{White/Caucasian} students, the most frequently cited role models were \textit{tech influencers}, followed by \textit{family members}, \textit{friends}, and \textit{practitioners}, with some students also mentioning \textit{pioneer scientists} and \textit{teachers}.

Among \textit{heterosexual} students, the most frequently cited role models were \textit{tech influencers}, followed by \textit{family members}, \textit{friends}, \textit{pioneer scientists}, \textit{practitioners}, and \textit{teachers}. This reflects a broad range of both personal and professional figures that continue to shape their engagement with the field. Notably, while \textit{family members} and \textit{friends} were key influencers in their initial decision to pursue SE, \textit{tech influencers} and \textit{pioneer scientists} emerged as prominent role models over time, indicating a shift towards more public and professional figures.

On the other hand, among the \textit{LGBTQIA+} students, those identifying as \textit{bisexual} cited \textit{tech influencers} as their primary role models, whereas earlier influences also included \textit{friends} and \textit{teachers}, showing a narrowing of focus to industry figures. For \textit{gay} students, \textit{tech influencers} remained a consistent source of both influence and long-term inspiration, though the presence of \textit{teachers} and \textit{family members}, which were initial influencers, disappeared as role models. Finally, \textit{transgender} participants cited both a \textit{tech influencer} and a \textit{teacher} as role models, aligning with their initial influences but expanding the scope of figures they look up to in the field. Table \ref{tab:rolemodels} summarizes these results.

\MyBox{\textbf{\emph{RQ3 -- Summary.}} \small Role models play a significant role in sustaining students' interest in SE. Tech influencers were the most commonly cited, followed by family members, friends, pioneer scientists, practitioners, and teachers. While 61\% of students reported no specific role models, those who did were inspired by both public figures and personal connections. Across genders, tech influencers were prominent, with male students also citing practitioners and family members and female students mentioning pioneer scientists, teachers, and family members. Ethnically, tech influencers were key for all groups, while Black/African American students also highlighted pioneer scientists. Heterosexual students primarily looked up to tech influencers and family members, while LGBTQIA+ students focused on tech influencers, with transgender students also citing teachers.}

\input{tables/rolemodel}

\section{Discussions}

Previous research has emphasized the importance of various factors in motivating students to pursue STEM careers, particularly the influence of teachers and learning environments. For instance, teachers play a significant role in fostering social motivation and long-term persistence in STEM fields~\cite{williams2019will}, while positive educational experiences can counter the decline in students' interest in science as they progress through their education~\cite{osborne2003attitudes}. Our findings align with this, as teachers were cited as role models and influencers, especially among female students, showing that mentorship in educational environments remains a critical factor in shaping career choices in SE. However, our study diverges slightly in that tech influencers and family members emerged as the most common role models and influencers across gender and ethnic groups, suggesting that public figures and personal relationships may play a more prominent role in SE than has been highlighted in previous STEM studies.


The literature also highlights the importance of role models in promoting diversity and inclusion in STEM fields, particularly for students from underrepresented groups~\cite{doherty2016role, susor2018role}. Historical figures and renowned scientists from marginalized backgrounds have been shown to provide students with tangible, relatable examples that inspire perseverance and career engagement~\cite{rosenthal2013pursuit}. In our study, pioneer scientists were cited as role models by some students, particularly males and female students, reflecting the lasting influence of historical figures. However, tech influencers significantly outpaced pioneer scientists in terms of their impact, suggesting that modern-day public figures and industry leaders are more relatable and influential for today’s students than historical figures, at least in the context of SE.

Finally, while prior research underscores the value of promoting diverse role models to create inclusive educational environments~\cite{gladstone2021role, karunanayake2004relationship}, our findings extend this conversation by showing how role models vary across gender, ethnicity, and sexual orientation. For example, tech influencers were consistently cited across all demographics, while family members played a larger role for heterosexual students, and teachers were more frequently mentioned by LGBTQIA+ students. This demonstrates that diverse sources of inspiration are critical to fostering a sense of belonging and encouraging persistence among underrepresented students in SE, contributing to broader efforts to diversify the workforce~\cite{nelson2022role}.

\subsection{Implications to Research}\label{SCM}
As far as we know, this is one of the first studies to specifically investigate how role models influence students in SE, a topic that has been extensively explored in fields such as healthcare and other STEM disciplines. While the role of role models in fostering diversity, resilience, and long-term engagement has been well-studied in these areas, our research brings this important discussion into SE, a field that has seen limited exploration of this aspect. By focusing on who motivates and inspires students to both enter and remain in SE, our study addresses a significant gap in the existing literature.

Our research is unique in its exploration of motivations, influences, and role models across several dimensions of diversity, including gender, ethnicity, and sexual orientation. By capturing these dimensions, we add to the growing body of knowledge about how different demographic groups experience their journey into SE education. Previous studies have examined STEM motivation broadly or focused on specific demographic groups, often without considering how multiple diversity dimensions intersect. In contrast, our study provides a nuanced view of how students from various backgrounds not only choose to enter the field but also remain engaged, which can inform future diversity and inclusion efforts in SE education research.

The implications of our findings for future research are significant and multifaceted. Our study opens the door for further investigation into the long-term effects of role models within SE, particularly the evolving impact of modern tech influencers compared to traditional pioneer scientists or historical figures. Additionally, studying how the roles of influences and role models shift as students progress from initial interest to professional careers could offer valuable insights into retention and engagement strategies. There are also opportunities to explore how role models may vary in impact based on other diversity dimensions, such as socioeconomic status, disability, or neurodiversity, which were not fully covered in our study. Understanding the traits and attributes of role models that resonate most with underrepresented groups could guide mentorship and outreach programs aimed at increasing diversity in the field. Finally, longitudinal studies tracking the evolution of students' motivations and role models over time would provide deeper insights into how these factors interact and shape career trajectories, ultimately contributing to a more diverse SE workforce.

\subsection{Implications to Educational Practice}
Our findings offer important implications for educational practice, particularly for SE professors who seek to foster a more inclusive and motivating learning environment. By recognizing the diverse motivations, influences, and role models that inspire students from underrepresented groups, professors can tailor their teaching strategies to support these students more effectively. For example, incorporating discussions about historical and contemporary figures in SE, especially those from underrepresented backgrounds, can help create a more relatable and inspiring learning experience. Highlighting the achievements of tech influencers, pioneer scientists, and practitioners can help students see tangible pathways for their own success in the field, while acknowledging the critical role of family members, friends, and teachers can help professors build supportive relationships that reflect these personal influences.

Additionally, educators can use these insights to create classroom activities that encourage students to explore and articulate their career motivations and aspirations. By integrating role models into course content—whether through case studies, guest lectures, or collaborative projects—professors can offer students more opportunities to engage with the figures who inspire them. Furthermore, mentorship programs and peer-to-peer learning activities can be designed to align with the diverse influences and role models students identify with, ensuring that all students, regardless of their gender, ethnicity, or sexual orientation, feel supported in their educational journey. These strategies not only help maintain students’ interest and engagement in SE but also contribute to building a more inclusive and diverse workforce for the future.

\subsection{Threats to Validity}
Our study has some limitations primarily related to selection and sampling bias, as we faced challenges in determining the global population of SE students. To address this, we used convenience sampling, snowball sampling, and crowdsourcing to expand our participant pool. Crowdsourcing was particularly useful in ensuring a more geographically diverse sample, aiming to correct the single-country bias from a prior study. Our efforts focused on capturing diversity in gender, ethnicity, and sexual orientation, which aligned with the study's objectives.

Another limitation concerns potential response bias, as participants from underrepresented groups may have felt inclined to provide socially desirable responses. Although we incorporated open-ended questions to minimize this effect and encourage unbiased reflections, the possibility remains. Lastly, while our findings might not be fully generalizable in a positivist sense, we believe they offer adaptable insights that could be reanalyzed in different contexts, contributing to valuable discussions for educators and researchers.

\section{Conclusion}

In this study, we surveyed 89 undergraduate students in SE and related fields to understand the complex interplay of motivations, influences, and role models that shape the career aspirations of students from underrepresented groups in SE. Our findings reveal that while career advancement, technological enthusiasm, and personal growth are significant motivators, these vary across gender, ethnicity, and sexual orientation. Influences such as family members, teachers, and tech influencers play critical roles in students’ decisions to pursue and remain in the field, with role models providing long-term inspiration. The lack of diversity in SE remains a pressing challenge, but understanding how these dynamics impact underrepresented students can help create more inclusive environments. In future work, we plan to create initiatives to address these groups' specific needs and experiences to create a more equitable SE landscape.

\section{Data Availability}
The data obtained from the survey is available at: \url{https://figshare.com/s/2592f96db738a1dc3c5a}.

\ifCLASSOPTIONcaptionsoff
  \newpage
\fi

\balance
\bibliographystyle{IEEEtran}
\bibliography{bib.bib}

\end{document}

%% file: tables/survey.tex
\begin{table}[h!]
\label{tab:survey}
\caption{Survey Questionnaire}
\centering
\normalsize
\begin{tabular}{p{8cm}}
\hline
\textbf{\newline ABOUT YOUR UNDERGRADUATE COURSE} \\\\

1. In which country is your university located? \\ \\

2. What is your course? \\ \\

3. What is your current year of study? \\ \\
\hline

\textbf{ \newline ABOUT YOUR DECISION TO TAKE A CAREER IN SOFTWARE ENGINEERING} \\\\

4. What motivated you to enroll in a technology program (e.g., computer science, software engineering)? \\ \\

5. Is there anyone you know, professional, scientist, or personality who influenced your decision to enroll in a technology program? Who? \\
\\

6. Do you have any role model that motivates you to be in a software engineering or related program? Who? \\ \\
\hline

\textbf{\newline ABOUT YOU} \\ \\

What ethnic group do you identify with? \\
(  ) White/Caucasian \\
(  ) Black/African American \\
(  ) Hispanic/Latinx \\
(  ) Asian/Pacific Islander \\
(  ) Native American/Indigenous \\
(  ) Middle Eastern/North African \\
(  ) Multiracial/Mixed Ethnicity \\
(  ) Other, which one?\\

\\
What best describes your gender?\\
(  ) Female \\
(  ) Male \\
(  ) Non-Binary \\
(  ) A gender not listed here \\
(  ) Prefer not to answer \\
\\

Do you identify as transgender? \\
(  ) Yes \\
(  ) No \\
\\

What is your sexual orientation? \\ \\
\hline
\end{tabular}
\end{table}

%% file: tables/Demographics.tex
\begin{table}
\centering
\caption{Demographics}
\renewcommand{\arraystretch}{1}
\label{tab:Demographics}
\begin{tabular}{llr}
\hline\noalign{\smallskip}
\multirow{3}{*}{ \textbf{Gender} } 
& Men & 64\\
& Women & 23$^a$ \\
& Non-Binary & 2 \\ \midrule 

\multirow{5}{*}{ \textbf{Ethnicity} } 
& White & 47 \\
& Black & 15 \\
& Hispanic & 13 \\
& Asian & 12 \\
& Mixed Ethnicity & 2 \\ \midrule 

\multirow{8}{*}{ \textbf{Sexual Orientation} } 
& Straight & 66 \\
& Bisexual & 10 \\
& Gay & 6 \\
& Lesbian & 2 \\
& Pansexual & 1 \\
& Asexual & 1 \\
& Other & 1 \\
& Prefer not to answer & 2 \\ \midrule 

\multirow{5}{*}{ \textbf{Program Focus} } 
& Computer Science & 55 \\
& Software Engineering & 10 \\
& Information Technology & 8 \\
& Computer Engineering & 7 \\
& Other Software-Related & 9 \\ \midrule 

\multirow{5}{*}{ \textbf{Year} } 
& 1st & 7 \\
& 2nd & 18 \\
& 3rd & 31 \\
& 4th & 19 \\
& 5th & 14 \\ \midrule 

\multirow{20}{*}{ \textbf{Country} }
& Australia & 3 \\
& Austria & 1 \\
& Brazil & 7 \\
& Canada & 7 \\
& Chile & 2 \\
& Estonia & 1 \\
& Germany & 1 \\
& Greece & 2 \\
& Hungary & 3 \\
& Israel & 1 \\
& Italy & 3 \\
& Latvia & 1 \\
& Mexico & 9 \\
& Netherlands & 4 \\
& New Zealand & 1 \\
& Poland & 10 \\
& Portugal & 13 \\
& South Africa & 11 \\
& United Kingdom & 2 \\
& United States & 7 \\

\noalign{\smallskip}\hline
\end{tabular}

\flushleft
\footnotesize{Notes: $^a$19 cisgender, 4 transgender}
\end{table}


%% file: tables/motivations.tex
\begin{table}[h!]
\centering
\label{tab:motivations}
\caption{Motivations for Students Pursuing Software Engineering}
\begin{tabular}{m{1.3cm} m{2cm} m{4cm}}
\hline
\textbf{Group} & \textbf{Category} & \textbf{Main Motivators} \\ \hline
\textbf{Sample} & All Participants & Career Advancement, Technological Enthusiasm, Personal Growth, Early Age Contact, Personal Interest, Previous Training, Social Contribution, Parental Influence \\ \hline
\textbf{Gender} & Male & Career Advancement, Technological Enthusiasm, Personal Growth, Personal Interest \\ \cline{2-3}
                & Female & Career Advancement, Personal Growth, Early Age Contact, Technological Enthusiasm \\ \cline{2-3}
                & Non-Binary & Career Advancement, Personal Interest \\ \hline
\textbf{Ethnicity} & Asian/Pacific Islander & Career Advancement, Personal Growth \\ \cline{2-3}
                   & Black/African American & Career Advancement, Technological Enthusiasm, Social Contribution \\ \cline{2-3}
                   & Hispanic/Latinx & Personal Growth, Technological Enthusiasm \\ \cline{2-3}
                   & Multiracial/Mixed Ethnicity & Career Advancement, Early Age Contact \\ \cline{2-3}
                   & White/Caucasian & Career Advancement, Technological Enthusiasm, Personal Growth \\ \hline
\textbf{Sexual Orientation} & Heterosexual & Career Advancement, Technological Enthusiasm, Personal Growth \\ \cline{2-3}
                            & LGBTQIA+ & Career Advancement, Technological Enthusiasm \\ \cline{2-3}
                            & Transgender & Career Advancement, Early Age Contact, Personal Growth, Technological Enthusiasm, Career Advancement \\ \hline
\end{tabular}
\end{table}

%% file: tables/influences.tex
\begin{table}[h!]
\centering
\label{tab:influences}
\caption{Summary of Influences on Students' Career Choices in Software Engineering}
\begin{tabular}{m{1.3cm} m{2cm} m{4cm}}
\hline
\textbf{Group} & \textbf{Category} & \textbf{Main Influences} \\ \hline
\textbf{Sample} & \textbf{All Participants} & Family Members, Tech Influencers, Teachers, Friends, Pioneer Scientists \\ \hline
\textbf{Gender} & \textbf{Male} & Tech Influencers, Friends \\ \cline{2-3}
                & \textbf{Female} & Family Members, Teachers \\ \cline{2-3}
                & \textbf{Non-binary} & No cited influences \\ \hline
\textbf{Ethnicity} & \textbf{Asian/Pacific Islander} & Friends \\ \cline{2-3}
                   & \textbf{Black/African American} & Tech Influencers, Pioneer Scientists, Family Members \\ \cline{2-3}
                   & \textbf{Hispanic/Latinx} & Tech Influencers, Teachers \\ \cline{2-3}
                   & \textbf{Multiracial/Mixed Ethnicity} & Family Members, Teachers \\ \cline{2-3}
                   & \textbf{White/Caucasian} & Family Members, Tech Influencers \\ \hline
\textbf{Sexual Orientation} & \textbf{Heterosexual} & Tech Influencers, Family Members \\ \cline{2-3}
                            & \textbf{Gay} & Teachers, Tech Influencers, Family Members \\ \cline{2-3}
                            & \textbf{Bisexual} & Friends \\ \cline{2-3}
                            & \textbf{Lesbian} & Teachers \\ \cline{2-3}
                            & \textbf{Transgender} & Family Members \\ \hline
\end{tabular}
\end{table}

%% file: tables/rolemodel.tex
\begin{table}[h!]
\centering
\caption{Role Models Across All Participants, Gender, Ethnicity, and Sexual Orientation}
\label{tab:rolemodels}
\begin{tabular}{m{1.3cm} m{2cm} m{4cm}}
\hline
\textbf{Group} & \textbf{Category} & \textbf{Main Role Models} \\ \hline
\textbf{Sample} & All Participants & Tech Influencers, Family Members, Friends, Pioneer Scientists, Practitioners, Teachers, No Role Model \\ \hline
\textbf{Gender} & Male & Tech Influencers, Family Members \\ \cline{2-3}
                & Female & Tech Influencers, Pioneer Scientists \\ \cline{2-3}
                & Non-Binary & Teacher \\ \hline
\textbf{Ethnicity} & Asian/Pacific Islander & Tech Influencers, Family Members \\ \cline{2-3}
                   & Black/African American & Tech Influencers, Pioneer Scientists \\ \cline{2-3}
                   & Hispanic/Latinx & Tech Influencers \\ \cline{2-3}
                   & Multiracial/Mixed Ethnicity & Teacher \\ \cline{2-3}
                   & White/Caucasian & Tech Influencers, Family Members, Friends, Practitioners \\ \hline
\textbf{Sexual Orientation} & Heterosexual & Tech Influencers, Family Members, Friends \\ \cline{2-3}
                            & Bisexual & Tech Influencers \\ \cline{2-3}
                            & Gay & Tech Influencers \\ \cline{2-3}
                            & Transgender & Tech Influencers, Teacher \\ \hline
\end{tabular}
\end{table}